\documentclass[%
 reprint,
 amsmath,amssymb,
 aps,
]{revtex4-1}

\usepackage{graphicx}
\usepackage{dcolumn}
\usepackage{bm}%

\begin{document}
\preprint{APS/123-QED}

\title{Hidden order behind two dimensional random vortices}

\author{Chien-chia Liu}
 \email{chien-chia.liu@ds.mpg.de}
 \affiliation{%
 Max Planck Institute for Dynamics and Self-Organization,\\
 Am Fa{\ss}berg 17, D-37077 G{\"o}ttingen, Germany
}%

\date{\today}

\begin{abstract}
The vortices that appear repeatedly and suggest turbulent dynamics are crucial to the understanding of sheared turbulence. These vortices produce order out of chaos, benefiting the turbulence modelling that focuses only on statistically stable quantities. In three dimensions, the hairpin vortices [J. Zhou, R. Adrian, S. Balachandar, T. Kendall, J. Fluid Mech., 387, 353 (1999); R. Adrian, Phys. Fluids, 19, 041301 (2007)] play such a fundamental role in the transport of momentum and energy for wall bounded sheared turbulence. In two dimensions where the atmospheric flow has a great play, the fundamental vortices have yet to be observed. Here we report the experimental finding of the `slanted' vortices in the quasi-two-dimensional sheared soap-film turbulence. We also demonstrate how these slanted fundamental vortices, which result from the interaction between the sheared mean flow and the random turbulent vortices, could possibly connect to the turbulent transport. The finding here might introduce another means of understanding the planetary turbulence where the mean flow is commonly existent and nonuniform.
\end{abstract}

\maketitle


\section{Introduction}
Knowledge of two-dimensional (2D) turbulent flow is important to the understanding of the quasi-2D flow surrounding our earth, with typical examples such as atmospheric and oceanic flows. However, 2D turbulence is quite different from their three-dimensional (3D) counterparts, due simply to the different nature between planar and volumetric flow dynamics. One remarkable contrast is the opposed sign in the eddy viscosity ($\nu_E$) \citep{K67,Starr}. By analogy with the inherent fluid viscosity ($\nu >0$) that dissipates energy from the mean flow, the activities of turbulent eddies in 3D that extract energy from the mean flow are equivalent to a virtual viscosity $\nu_E>0$ \citep{TL}. Therefore, sheared turbulent flow requires higher driven power than sheared laminar flow in order to overcome the higher total viscosity, i.e. the sum of $\nu$ and $\nu_E$. However, the mechanism of generating negative $\nu_E$ is still much less understood than that of creating positive $\nu_E$ \citep{K67,Starr,TL}. In view of the important role of hairpin vortices \citep{Hairpin1,Hairpin2} in 3D sheared turbulence, we can thus expect the existence in 2D of some kind of fundamental vortices that are responsible for negative $\nu_E$. In the sense of fundamentality, such vortices should be repeatedly identifiable in the flow domain, and should clearly imply negative $\nu_E$. Here we experimentally realize the phenomenon of negative eddy viscosity, $\nu_E<0$, in quasi-2D sheared turbulence, and demonstrate a possible mechanism of such phenomenon by revealing the hidden `slanted' fundamental vortices in 2D. Here we adopt the conventional definition of eddy viscosity which reads \citep{Starr,TL,S87,CH}
\begin{equation}
{\nu _E} \equiv  - \overline {uv} {\left( {\frac{{\partial U}}{{\partial y}}} \right)^{ - 1}},
\end{equation}
where $u$ and $v$ are respectively the streamwise ($x$) and the wall-normal ($y$) turbulent fluctuating velocities, $\overline{( )}$ is the time average, $-\overline {uv}>0$ in 3D \citep{TL} is the cross-correlation of $u$ and $v$ or simply the turbulent shear stress, $U$ is the streamwise mean velocity, and $\partial U/\partial y$ is the shear strain or velocity gradient of the mean flow. For instance, the consideration of $\partial U/\partial y>0$ allows us to denote that $\nu_E<0$ corresponds to $-\overline {uv}<0$ while $\nu_E>0$ is associated with $-\overline {uv}>0$.

\begin{figure}
\centerline{
\includegraphics[width=0.45\textwidth, clip, trim=120 100 250 160, angle=0]{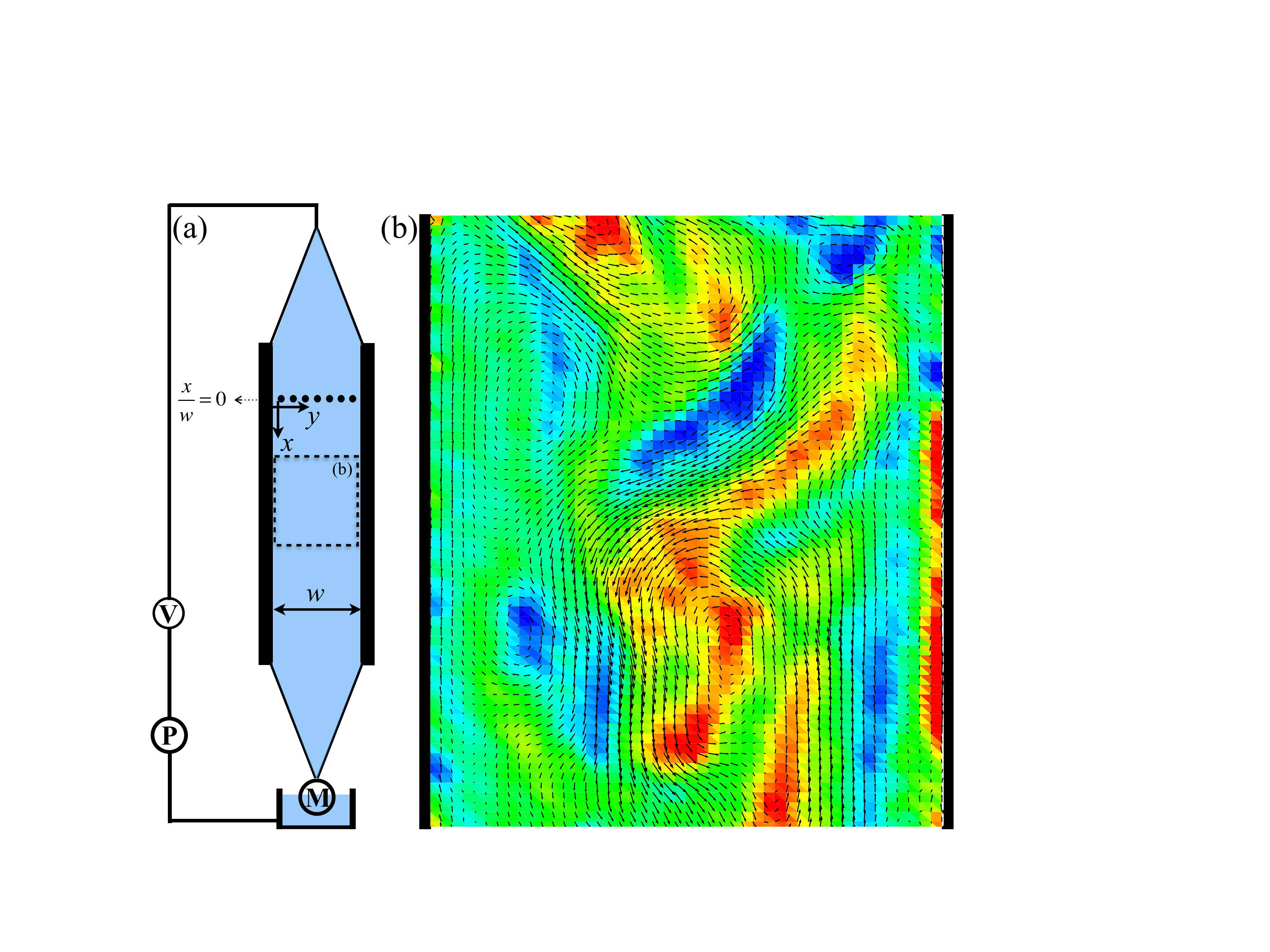}
}
\caption{\label{fig:epsart} 
(a) Typical setup of the soap-film channel for quasi-2D sheared flow. The film is framed by two parallel thin metal blades with a separation $w$, in addition to the entrance and exit sections formed by the fishing wires. The exit section is stabilized by a mass $M$. The soap solution is circulated by the pump $P$ and its flow rate is regulated with a valve $V$. The streamwise axis $x$ lies on the top of the left channel border, with the wall-normal axis $y$ set by the right-hand rule. (b) The measuring domain reveals the flow structures downstream of a comb ($x/w=0$) via the superimposed field of fluctuating velocities and associated fluctuating vorticities, where the domain centres at $x/w \approx 7.6$. Red/blue spots correspond to positive/negative vorticities.
}
\end{figure}

\begin{figure*}
\centerline{
\includegraphics[width=0.85\textwidth, clip, trim=0 180 0 120, angle=0]{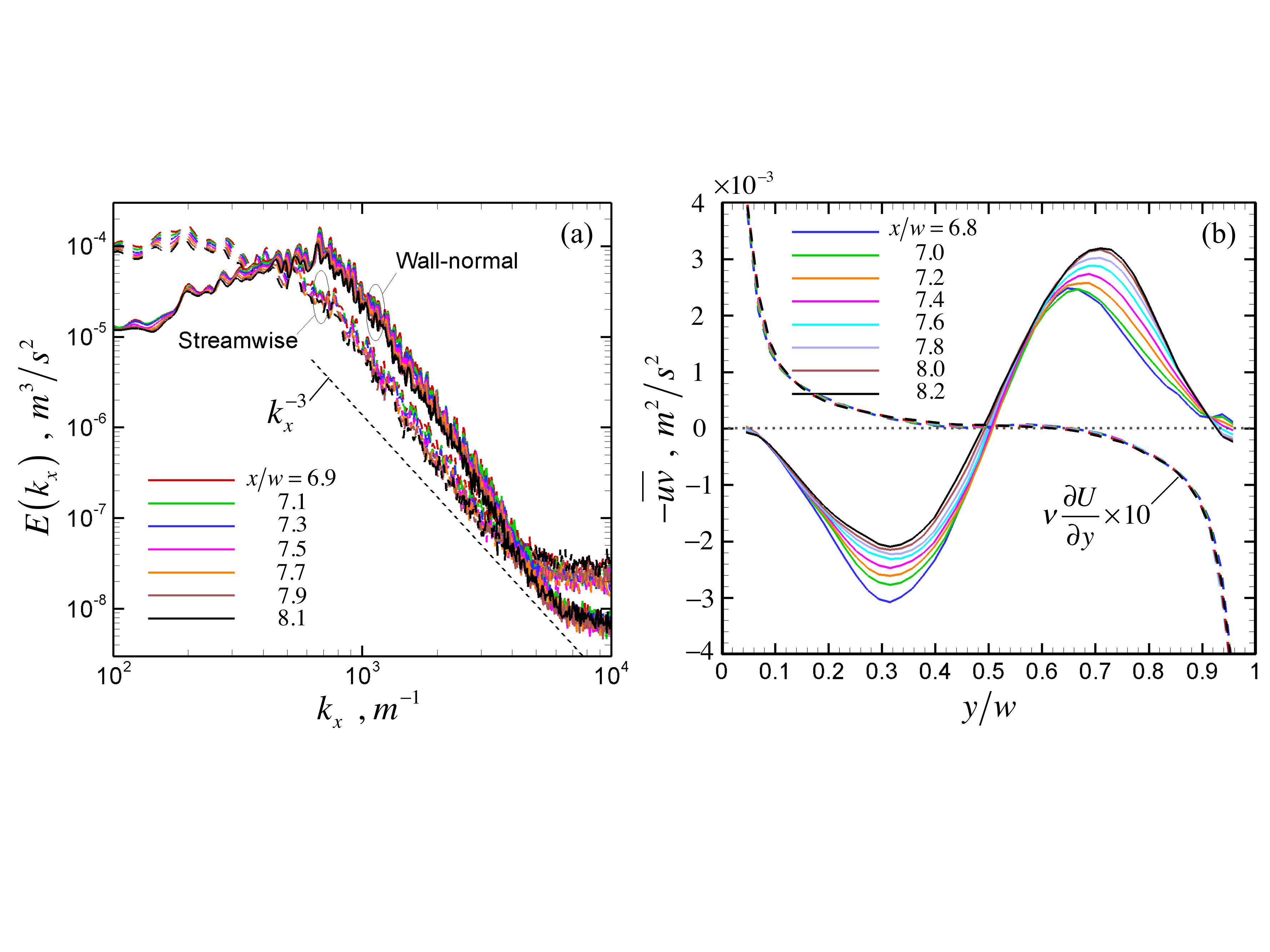}
}
\caption{\label{fig:epsart} 
(a) Kinetic energy spectra of both streamwise and wall-normal fluctuating velocity components at several downstream positions, revealing the prevalence of enstrophy-like cascade in the current sheared turbulence. (b) Distributions of turbulent shear stress (solid lines) and viscous shear stress (dashed lines) at several downstream positions. Note that in the range of $y/w$ from 0 to 0.5, the viscous shear stress is positive and thus $\partial U/\partial y>0$ while the turbulent shear stress ($-\overline {uv}$) is negative.
}
\end{figure*}

Experiment demonstration of the statistical negativeness of $\nu_E$ is relatively lacking \citep{Tsinober,TsinoberText}, even though it has been theoretically confirmed to be possible in 2D weakly perturbed flows \citep{Gama} and in turbulent flows \citep{Starr,K76}, and has also received support from direct numerical simulation \citep{S87,CH}. Here we perform the classical experiments of quasi-2D flow with the soap-film channel (Fig. 1a) \citep{Soap1,Soap2,Walter} and determine the turbulent fluctuating fields of velocities and vorticities (Figs. 1b,3a) with particle image velocimetry (PIV). We aim to show the existence of `slanted' fundamental vortices in quasi-2D sheared turbulence which displays the enstrophy-like forward cascade, and to demonstrate the contribution of these vortices to statistically negative $\nu_E$. Note while in one case where there is a mean flow negative $\nu_E$ is likely to create merely inverse energy transport in which the energy flow goes from turbulence to mean flow \citep{Starr}, in the other without a mean flow negative $\nu_E$ is prone to establish the inverse energy cascade because the energy flows from small-scale turbulence to large-scale turbulence \citep{K76}. The present study belongs to the former without presenting the inverse energy cascade, which has also been confirmed in direct numerical simulations \citep{S87,CH}.
 
\section{Experiments}
The current vertical soap-film channel (Fig. 1a) has a parallel section of about 1.6 m. The stainless steel blades of 0.5 mm in thickness are used as the parallel boundaries. The convergent and divergent sections are formed with the fishing wire of 0.5 mm in diameter. The lengths are respectively 0.6 m and 0.2 m for the divergent and convergent sections. Approximately uniform film thickness of the order of 10 $\mu$m, except in the neighborhood to the wall, has been identified in a similar setting of soap-film channel \citep{Soap1,Soap2,Tran2010N,Janus}. Several different kinds of pumps with or without the overhead solution reservoir have been tested in the same disturbed flow conditions, and no apparent differences in their associated wavenumber spectra is observed. The influence of the low-wavenumber pulses or modulations from the pump on the turbulent flow of our current interest is thus negligible. A comb with equally-spaced thin pins is placed somewhere downstream of the entrance of the parallel channel (Fig. 1a) in order to create nearly homogeneous turbulence with downscale enstrophy-like cascade in a sizeable domain excluding the neighborhood to the wall where the effect of mean shear is not negligible. For the selected case shown here, the channel width ($w$) is 23 mm and both the diameter of thin pins and the gap between adjacent pins are about 0.7 mm. Note that the presented results are not sensitive to either sizes of uniform gaps or uniform pins of the comb. The flow Reynolds number, $Re = \overline U w/2\nu \approx 21850$ for the case demonstrated here, where the profile mean velocity $\overline U = \int_0^w {U(y)dy}/w \approx 1.9$ m/s at the midstream of the current flow domain, and $\nu$ is the kinematic viscosity of water at $20^\circ$C.

\begin{figure*}
\centerline{
\includegraphics[width=0.8\textwidth, clip, trim=50 100 50 20, angle=0]{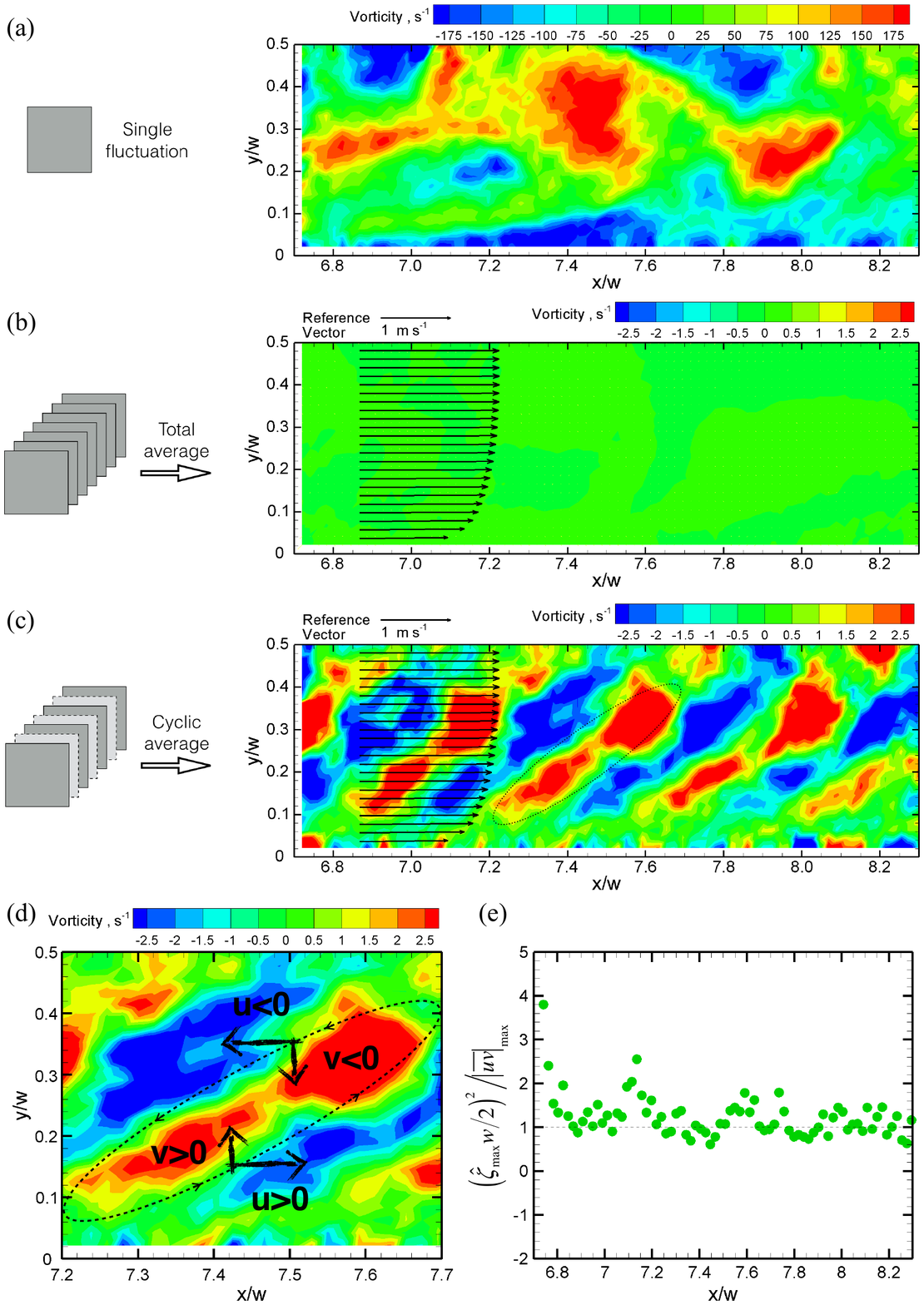}
}
\caption{\label{fig:epsart} 
(a) An instantaneous frame of the fluctuating vorticity field. (b) The total average result using all frames in the time series, showing the mandatory zero-mean of any fluctuating signals. (c) The cyclic average result according to Eq. (2) using the frames sampled at a lower yet `right' frequency, revealing the `slanted' fundamental vortices in the flow domain whose $\partial U/\partial y \geq 0$ as shown by the superimposed vector profile. (d) A close-up view of a slanted vortex as identified in (c) by a broken oval. The arrows illustrate the flow directions associated with the vortex circulation. (e) The ratio of the maximum stress of the slanted vortices to the maximum turbulent shear stress, as a function of the streamwise positions.
}
\end{figure*}

PIV measurements are performed with 10-$\mu$m seeding particles in the soap solution of $\sim$2\% volume/volume percent concentration of the commercial detergent. The PIV particles in the soap film are illuminated, thanks to the quasi-2D film geometry, by projecting high luminance spot lights onto the film's area of interest. There has to be an angle between the incident light axis and the film's normal in order to create sufficient light scattering by particles, and also to avoid directly exposing the camera sensor to strong incident light. All measurements in this study are performed in the stable flow regime, well above where the Marangoni instability resides \citep{Marangoni}. The PIV images used here are taken at the frame rate of 10 kHz with 1,280 $\times$ 800 pixel$^2$. The resulting resolution is about 0.029 mm/pixel. To calculate the velocity vector field, the size of the square correlation window is set as 64 pixel with a window overlap of $75\%$. Instantaneous vorticities in a velocity field are then calculated with a central difference method. The typical field of fluctuating velocities after subtracting the mean is given in Fig. 1(b), with the superposition of its associated fluctuating vorticity field.

\section{Results}
The variations of streamwise and wall-normal turbulent kinetic energy spectra $E(k_x)$ in terms of different streamwise wavenumbers $k_x$ (Fig. 2a) are obtained at different positions along the centerline of the soap-film channel from $x/w \approx 7$ to 8 downstream of the comb at $x/w = 0$. The length scales are calculated by transforming via Taylor's hypothesis the local Eulerian velocity information varying with time into the spatial velocity information varying with the downstream distance \citep{TL}. It shows a similar $k_x^{\alpha}$ scaling where $\alpha \approx -3$ in the spectra regardless of position and direction, indicating the prevalence of downscale enstrophy-like cascade \citep{K67}. Note that decaying turbulence with scales of the enstrophy range in the soap-film channel has values of $\alpha = -3.3 \pm 0.3$ \citep{Walter}, suggesting steeper spectra than the theoretical prediction ($\alpha = -3$). The same results are found at the lines parallel to the channel centerline from $y/w \approx 0.1$ to 0.9, revealing the quasi-features of cascade isotropy and cascade homogeneity in the current domain of quasi-2D sheared turbulence. Isotropy of turbulent cascade is simply identified by the similarity in both streamwise and wall-normal energy spectra or equivalently by the same unambiguous spectral exponent $\alpha$. Homogeneity of turbulent cascade means that the cascade isotropy persists in the domain of measurement except near the walls. This domain of interest roughly covers the range of $y/w$ from 0.1 to 0.9 across the channel, as well as the range of $x/w$ from 4 up to more than 10 downstream of the comb. The interested reader may also find quite a few possible contributions to spectra with $\alpha < -3$, such as sharp vorticity gradients \citep{VorDiscont,SharpVorGrad}, band-limited forcing \citep{BandLimitForce}, and self-similar vortices \citep{Yakhot,Borue}.

To identify the sign of $\nu_E$ via Eq. (1), we need to know the sign of the viscous shear stress $\nu \partial U/\partial y$ as well as the turbulent shear stress $-\overline {uv}$. The profiles of $\nu \partial U/\partial y$ and those of $-\overline {uv}$ (Fig. 2b)  in terms of $y/w$ are shown at eight equally spaced positions from $x/w=6.8$ to 8.2. As we can see from Fig. 2(b), negative $\nu_E$ is evidenced in the sense of Eq. (1), except at the channel centerline and in proximity to the walls where $-\overline {uv}$ approaches zero. Note that the asymmetry of $-\overline {uv}$ profiles is owing to the imperfect upstream boundary conditions of the soap-film channel.

Is there a tentative explanation how negative $\nu_E$ is created based on 2D vortex structures? There seems to be if we propose a mechanism based on the slanted vortices (SV) which imply straightforwardly $-\overline {uv} < 0$ as can be seen in Figs. 3(c,d). We could expect that the vorticity intensity of SV is roughly identical in a fixed location of observation. However, SV must alternate the sign of its vorticity in any fixed locations so that no net vorticity is generated. Therefore, SV resulted from shear-turbulence interaction, if exists, should be able to be visualized (Figs. 3c,d) by cyclically averaging the fluctuating vorticity fields with the following relation:
\begin{eqnarray}
\zeta \left( {x,y} \right) = \frac{1}{{{n_{\max }} + 1}}\sum\limits_{n = 0,1,2,...}^{{n_{\max }}} {\xi \left( {x,y,{t_0} + n{\tau}} \right)},
\end{eqnarray}
where $\zeta$ is the vorticity field of SV, $\xi$ is the fluctuating vorticity field, $n=\{0,1,2,...,n_{\max}\}$, $n_{\max}$ is the number of instantaneous fields available for the conditional average ($=1092$ here), $t_0$ is an arbitrary time origin ($=0$ here), and $\tau$ is the time scale which characterize the transport of SV in the mean flow. Incidentally, the concept of cyclic average in Eq. (2) is basically identical to what Hussain has called it `phase average' more than thirty years back \citep{Hussain}. It also deserves a note that SV cannot exist in the total average field (Fig. 3b) which is the time average of all fluctuating vorticity fields, because of the periodic nature of the fluctuating signals.

Because of the no-slip condition ($U=0$) at the wall leading to the sheared mean flow, the transport of SV in the sheared mean flow is nonuniform. It is very likely such nonuniform transport that tilts SV, rendering the feature of `slant' to the SV here. Therefore, the time scale of mean shear strain ($\tau_S$) is used to define $\tau \equiv a\tau_S = a\delta/\overline U$, where $a$ is supposed to be some constant of the order 1, and $\delta$ is the velocity boundary-layer thickness which takes a constant of $w/2$ here. Figure 3(c) presents the clearest visualisation of SV we can find here, with the choice of $a=0.82$ after trial and error, whose order of magnitude also implies the presumed importance of $\tau_S$. A close-up view of a selected SV enclosed by a broken oval in Fig. 3(c) is also shown (Fig. 3d) for clarity. It is clear in Fig. 3(d) that SV is strained and tilted by the sheared mean flow where $\partial U/\partial y > 0$, so that the major axis of the oval eddy roughly leans along the mean velocity gradient. Such remarkable geometry and orientation of SV implies that $-\overline {uv}<0$, because most of the surrounding flow parallels the major axis of the oval SV, as illustrated in Fig. 3(d) with the direction-guided arrows. Note that $\tau_S$ in Eq. (2) not only implies the important role of sheared mean flow in the interaction with turbulence, but also roughly represents the time scale of the transport of SV by the nonuniform mean flow.

Since SV can manipulate the sign of $-\overline {uv}$, we are then curious to glimpse whether SV could possibly regulate the intensity of $-\overline {uv}$. With the length scale $\delta$ and the time scale $\zeta^{-1}$ of SV in Figs. 3(c,d), we may roughly estimate the intensity of SV. A simple quantitative estimation (Fig. 3e) that constructs a bridge between the peak intensity of SV and the peak intensity of $-\overline {uv}$ at a fixed streamwise location can be written as:
\begin{eqnarray}
{\left[ {{{\hat \zeta }_{\max }}\left( x \right)\delta} \right]^2} \approx {\left| {\overline {uv} } \right|_{\max }}\left( x \right),
\end{eqnarray}
where $\hat \zeta_{\max}$ is the maximum vorticity of SV acquired via Eq. (2) with all possible $t_0$, and $\left| {\overline {uv}} \right|_{\max}$ is the absolute peak value in a $-\overline {uv}$ profile. The estimations in Fig. 3(e) give reasonable supports to Eq. (3), further implying that the behavior of $-\overline {uv}$ here could also be regulated by the SV.

\begin{figure}
\centerline{
\includegraphics[width=0.5\textwidth, clip, trim=125 240 145 190, angle=0]{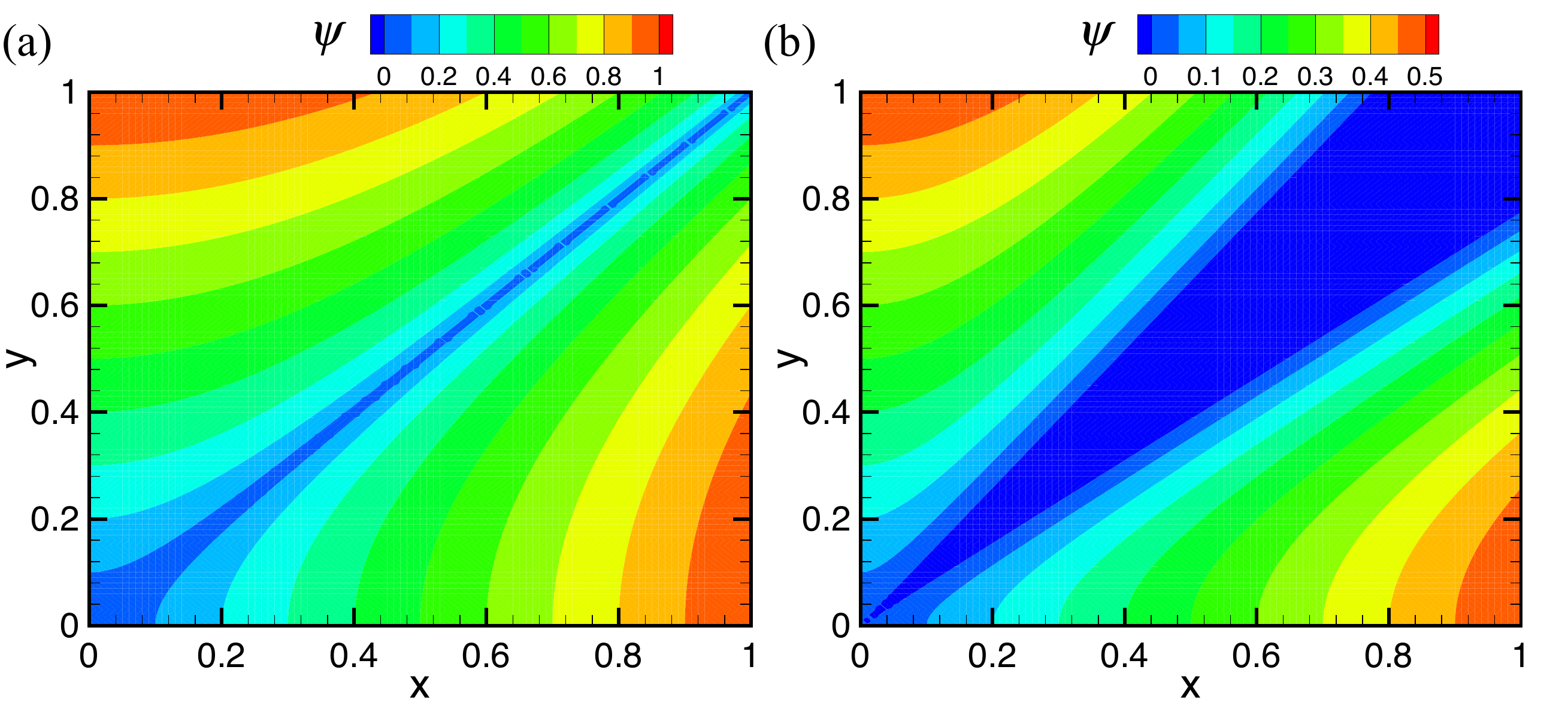}
}
\caption{\label{fig:epsart} 
(a) Stream functions of $\psi = \Re \left\{ {\sqrt { \pm {x^2} \mp {y^2}} } \right\}$. (b) $\psi  = \Re \left\{ {\sqrt { \pm {x^2} \mp {y^2}} } \right\} - 0.5 \sqrt {{x^2} + {y^2}}$.
}
\end{figure}

\section{Discussion}
To see more about how SV is possibly formed in 2D sheared turbulence, we turn to consider the following disturbance-energy equation \citep{S87} for inviscid flow 
\begin{eqnarray}
\frac{\partial }{{\partial t}}\left\langle {\frac{1}{2}{{\left| {\nabla \psi } \right|}^2}} \right\rangle  =  - \left\langle {uv} \right\rangle \frac{{dU}}{{dy}},
\end{eqnarray}
where $\left\langle {} \right\rangle$ is the ensemble average, $\nabla$ is the gradient operator, and $\psi = \psi ( x,y,t)$ is the stream function for turbulent fluctuations. Here $u =  - \partial \psi/\partial y$ and $v =  \partial \psi/\partial x$ and thus $\xi  = {\nabla ^2}\psi$. Note that the product of $dU/dy$ and $\left\langle {uv} \right\rangle$ is in general positive (Fig. 2b). To consider a simple shear flow would be not only the least complicated but also meaningful enough. Therefore, with $dU/dy$ and $\left\langle {uv} \right\rangle$ both being constants, we may integrate Eq. (4) to have 
\begin{eqnarray}
\left\langle {\frac{1}{2}{{\left| {\nabla \psi } \right|}^2}} \right\rangle  =  - \left\langle {uv} \right\rangle \frac{{dU}}{{dy}}t + c,
\end{eqnarray}
where $c$ is a constant of integration. Assume that SV exists only for $t>0$ and has a lifetime cycle equivalent to $(dU/dy)^{-1}$. Thus, from Eq. (5) and $\psi {\nabla ^2}\psi  = \nabla  \cdot \left( {\psi \nabla \psi } \right) - {\left| {\nabla \psi } \right|^2}$, we may write down a tentative relation for `SV' at any positive integers of $t/(dU/dy)^{-1}$ (mimicking the cyclic average) 
\begin{eqnarray}
\left\langle {\psi {\nabla ^2}\psi } \right\rangle_{cyc}  = \left\langle {\psi \zeta } \right\rangle_{cyc}  = C < 0,
\end{eqnarray}
where the constant $C$ is negative here because of $-\left\langle {uv} \right\rangle_{cyc}<0$, and the subscript `cyc' means the cyclic average.

It then allows us to discuss the heuristic basic solutions to Eq. (6), which are $\psi  = \sqrt { \pm {x^2} \mp {y^2}}$. The real part of the aforesaid solutions, $\Re \left\{ {\sqrt { \pm {x^2} \mp {y^2}} } \right\}$, is depicted in Fig. 4(a). It clearly shows in Fig. 4(a) that the basic flow for Eq. (6) is signified by a $+45^\circ$ shearing, consistent with the orientation of SV found here (Figs. 3c,d). Furthermore, we may consider in Eq. (6) that $C<0$ is a statistical fact owing to $-\left\langle {uv} \right\rangle<0$. That is, we need as well to consider the influence of the less possible case of $C>0$ in Eq. (6), where its basic solution ($\psi  = \sqrt {{x^2} + {y^2}}$) tends to create isotropic vortices. Therefore, we may consider the basic flow for Eq. (6) with a more general form of $\psi  = k\Re \left\{ {\sqrt { \pm {x^2} \mp {y^2}} } \right\} - m\sqrt {{x^2} + {y^2}}$, where the positive constants $k>m$ owing to the greater weighting for the circumstances with $C<0$ than those with $C>0$. The result (Fig. 4b), with $k=1$ and $m=0.5$ for example, further diverges the slanted shearing, confirming the finding in the current study, where SV has a wider upper body than its lower part (Figs. 3c,d).

\begin{figure}
\hspace{-5 pt}
\centerline{
\includegraphics[width=0.4\textwidth, clip, trim=250 0 260 0, angle=0]{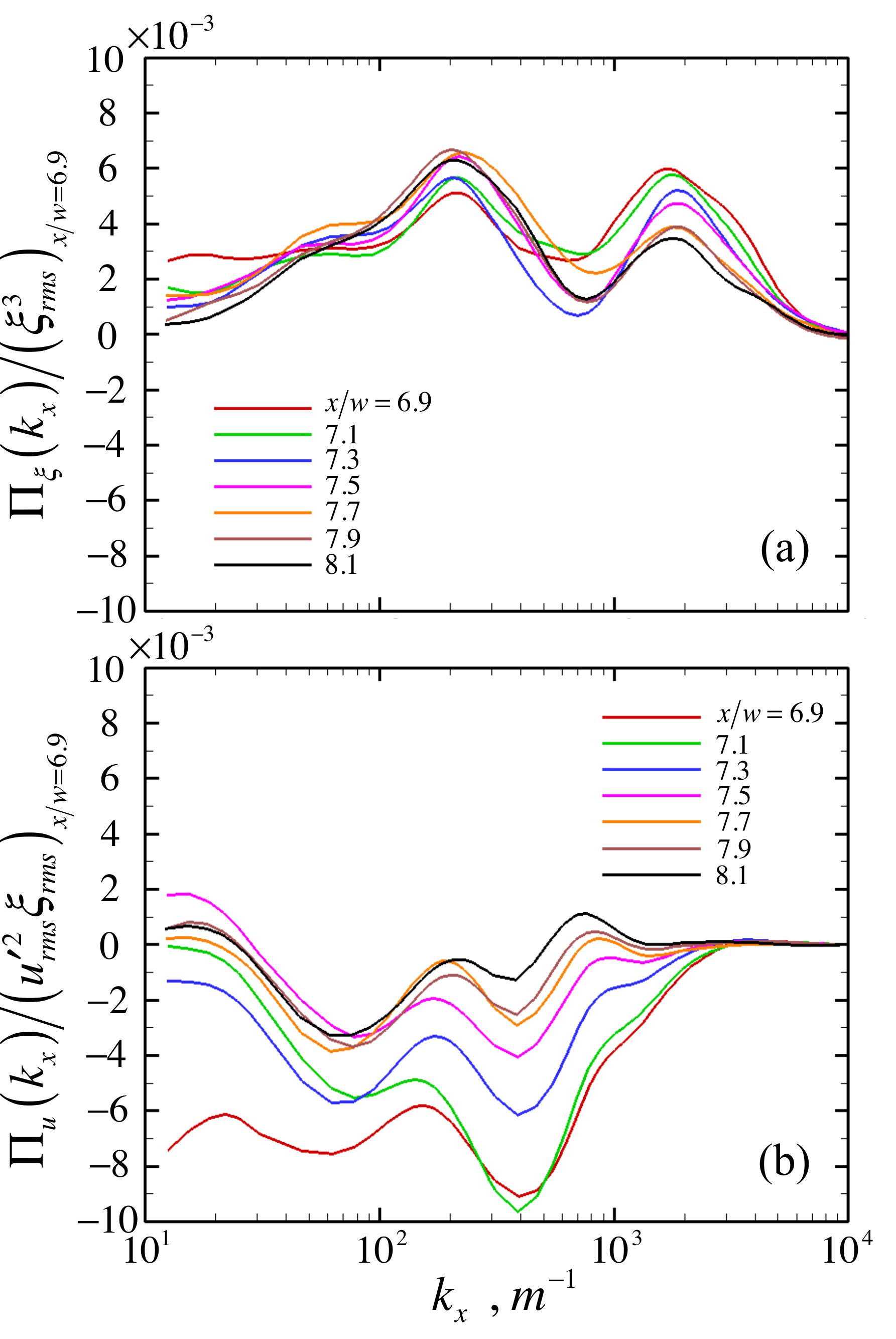}
}
\caption{\label{fig:epsart} 
(a) Enstrophy and (b) energy fluxes of the current case, normalized using the same upstream velocity and vorticity intensities.
}
\end{figure}

\begin{figure*}[htb]
\centerline{
\includegraphics[width=1\textwidth, clip, trim=30 350 40 0, angle=0]{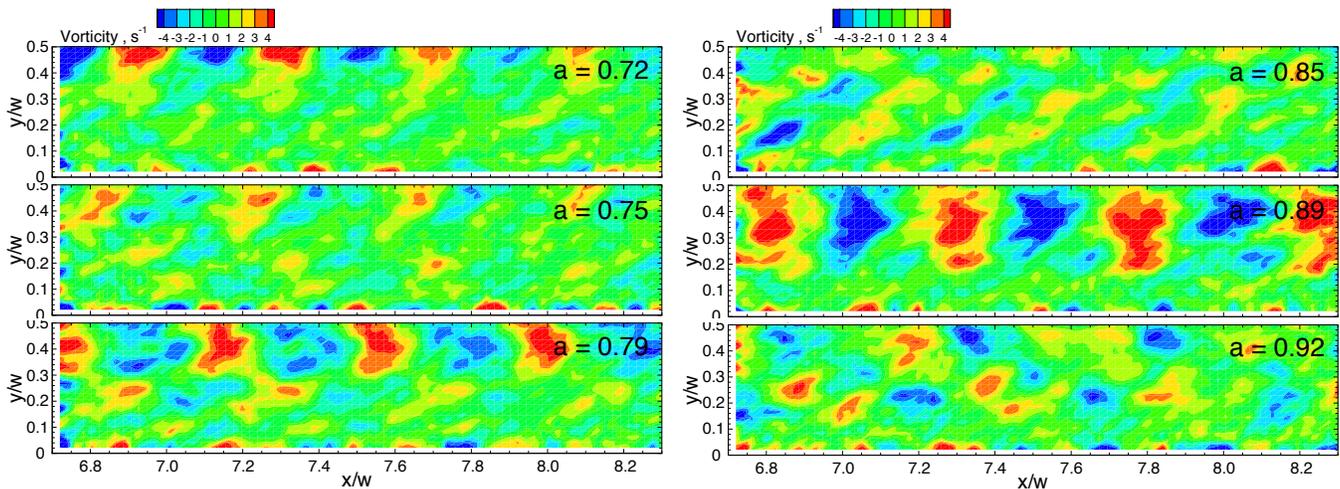}
}
\caption{\label{fig:epsart} 
The non-slanted vortex structures revealed by the cyclic average with different values of `$a$' in $\tau$.
}
\end{figure*}

We further show that the implication of Eq. (6) might not depend on $Re$. From Eq. (1) and Eq. (6), we have $\left\langle{\psi \zeta} \right\rangle \sim -\left\langle {uv} \right\rangle_{cyc} \sim \nu_E dU/dy$. Since here $dU/dy$ is considered as the corresponding mean shear (a constant) for any $Re$, we may have $\left\langle{\psi \zeta} \right\rangle/\overline U^2 \sim (\nu_E/\nu)(\nu dU/dy)/\overline U^2 \sim 1$, where $\nu_{E}/\nu \sim Re$ and $(\nu dU/dy)/\overline U^2 \sim 1/Re$.

The size of SV is roughly equal to $w/2$ which is one order of magnitude larger than the initial forcing scale of around 1 mm (either the spacing between the pins or the size of pins of the comb used here). The presence of the SV train thus implies not only the existence of inverse energy transport from small scales to large scales, but also the quasi-inviscid self-preserving nature of SV. To understand this, we estimate the enstrophy flux $\Pi_\xi$ (Fig. 5a) and energy flux $\Pi_u$ (Fig. 5b) of the current case, using the scale-by-scale energy budget equation with low-pass filtering \citep{Frisch}. These fluxes are determined at the channel centerline with the Taylor hypothesis, and then normalized with root-mean-square velocity $u_{rms}'=[(\overline {u^2}+\overline {v^2})/2]^{1/2}$ and vorticity $\xi_{rms}=(\overline {\xi^2})^{1/2}$ at the very upstream position for comparison. First, we see in Fig. 5(a) the positive (downscale) $\Pi_\xi$ in roughly the same enstrophy range of scales where $k_x > 10^3$ as in spectra of Fig. 2(a). $\Pi_\xi$ then seems to be completely dissipated at smaller scales around $k_x=10^4$. We also observe in Fig. 5(b) the negative (upscale) $\Pi_u$ from a similar forcing at about $k_x = 10^3$ to larger scales. However, $\Pi_u$ seems not to converge to zero at lower $k_x$, implying the absence of a dominant large-scale dissipation. More surprisingly, downscale $\Pi_\xi$ is found in the same range where upscale $\Pi_u$ exists. This suggests, seemingly, that the existence of SV or equivalently negative $\nu_E$ might be a result of flow structures self-organizing so as to reduce energy loss and to maximize their own lifetime.

At this point, the current study is still in the preliminary stage and many unknowns remain to be investigated. One particular issue is the systematic detection of important fundamental vortices. Owing to the periodic nature of the fluctuating signals, many possibilities of non-slanted vortices can be found using Eq. (2) with some arbitrary choices of $\tau$ (Fig. 6). The non-slanted vortices have not been taken into current consideration because they seem hardly to contribute constantly to the nonzero turbulent shear stress based simply on their geometries (Fig. 6). However, all possible modes that can be revealed by the cyclic average may help to understand a broader picture of 2D sheared turbulence. Hopefully, in the future studies this could render us a means of detailed discussion on the corresponding contribution of each mode in 2D sheared turbulence. Moreover, it is also interesting to know how the evolution of SV might lead to the Janus spectra where $\alpha = -3$ and $-5/3$ coexist in two perpendicular directions further downstream of the soap-film channel.

The importance of SV and thus negative $\nu_E$ might be pointed out in some familiar planetary circulations as well as severe weather events. For example, the current study might suggest the possible existence of a quasi-2D interaction between typhoon and turbulence in the atmosphere. A typhoon which is a vortex-like object from the view point of a satellite could create quasi-2D sheared mean flow \citep{Will}. Note the upper outflow shield of a typhoon revealed via cirrus clouds can easily be greater than, for example, the longitudinal span of Japan ($\sim$1000 km), and that the altitudes at which such shield resides range roughly from 10 to 15 km \citep{Will}. The quasi-2D argument can thus be reasonable because of the extremely large ratio of the vortex diameter to its thickness ($\sim$5 km). On the other hand, enstrophy cascade is known to exist in the atmosphere \citep{NG1984,Lindborg2011,Xia}. The atmospheric spectrum which is obtained at altitudes of around 9-13 km \citep{NG1984} is of double cascade. There exists an intermediate length scale separating two cascades of the order of about 100 km. Enstrophy cascade prevails at scales larger than the intermediate scale, while energy cascade occupies at smaller scales. In spite of the debates on the direction of such energy cascade or equivalently on the dimensionality of its associated turbulence, there is a wide agreement of correspondence between enstrophy cascade and quasi-2D turbulence \citep{Lindborg2011,Xia}. That whether the typhoon after landfall can possibly absorb energy from the atmosphere might be immediately a question to think about. Is it possible to identify any responsible structures in such flow?

\section{Preliminary conclusion}
We have experimentally studied quasi-two-dimensional sheared turbulence in the soap-film channel with smooth walls. Such turbulence is shown to have both quasi-enstrophy cascade and negative eddy viscosity via flow visualisation carried out by particle image velocimetry. We demonstrate the existence of slanted fundamental vortices by cyclically averaging the fluctuating fields. These slanted vortices show a promising connection to the phenomenon of negative eddy viscosity. We hope that the study of the identifiable fundamental vortices might open up another means to enhance our understanding of two-dimensional sheared turbulence.\\

Discussions with Rory T. Cerbus, Yuna Hattori, Pinaki Chakraborty and Tapan Sabuwala are gratefully acknowledged. This work was supported by the Okinawa Institute of Science and Technology Graduate University. The author is also grateful to the Max Planck Society for their support.

\providecommand{\noopsort}[1]{}\providecommand{\singleletter}[1]{#1}%

\end{document}